# Interplay among superconductivity, pseudogap, and stripe correlations in different high-$T_c$ cuprates

S H Naqib and R S Islam

*Department of Physics, Rajshahi University, Rajshahi-6205, Bangladesh*

E-mail: salehnaqib@yahoo.com

## Abstract

The effect of Zn substitution in the $CuO_2$ plane on the superconducting transition temperature, $T_c$, was studied for the $La_{2-x}Sr_xCu_{1-y}Zn_yO_4$ and $YBa_2(Cu_{1-y}Zn_y)_3O_{7-\delta}$ compounds over a wide range of hole concentration, $p$, and Zn content ($y$). Zn induced rate of suppression of $T_c$, $dT_c(p)/dy$, was found to be strongly $p$-dependent and showed a monotonic variation with $p$, except in the vicinity of $p \sim 0.125$, *i.e.*, near the so-called $1/8^{th}$ anomaly where the charge/spin stripe correlations are at their strongest in hole doped cuprates. The magnitude of $dT_c(p)/dy$ decreased significantly around this hole concentration implying that Zn suddenly became less effective in degrading $T_c$ near the $1/8^{th}$ anomaly for $La_{2-x}Sr_xCu_{1-y}Zn_yO_4$. The same feature, somewhat at a reduced scale, was also observed for $YBa_2(Cu_{1-y}Zn_y)_3O_{7-\delta}$ compounds. This is counterintuitive since static stripe order itself degrades superconducting order. We have discussed the possible scenarios that can give rise to such a non-monotonic $dT_c(p)/dy$ near $p \sim 0.125$. We have also looked at the $p$-dependent characteristic pseudogap energy scale, $\varepsilon_g(p)$, which shows a nearly linear decrease with increasing $p$ without any noticeable feature at $p \sim 0.125$. Moreover, there is no significant effect of Zn content on the value of $\varepsilon_g(p)$. All these observations are indicative of a complex and possibly competing interplay among the superconducting, pseudogap, and stripe correlations in the hole doped cuprates.

# 1. Introduction

Ever since the discovery of high-$T_c$ cuprates, there has been intense debate on the question of how the parent Mott antiferromagnetic (AFM) insulator transforms itself in to a high-$T_c$ superconductor upon carrier doping but the answer remains unclear, till date. The situation is complicated because these remarkable materials exhibit various normal state correlations (which may well exist in the superconducting state) as the number of added holes, *p*, in the CuO$_2$ plane is varied. Besides superconductivity itself, the most extensively studied phenomena are undoubtedly the pseudogap (PG) and the charge/spin stripe states [1 – 5]. The relevance of these correlations to the emergence of superconductivity is a matter of extensive study and no clear picture has emerged so far. It is generally thought that understanding of these normal state correlations will unlock the mystery of high-$T_c$ superconductivity in cuprates.

The PG is detected experimentally in the *T-p phase diagram* over a certain range of hole content, extending from the underdoped (UD) to the slightly overdoped (OD) regions. In the PG region a number of non-Fermi-liquid like features are observed both in normal and superconducting (SC) states where contrary to one of the central tenants of the Landau quasi-particle (QP) picture, low-energy excitations are gapped along certain directions of the Brillouin zone while Fermi arcs survive in other directions [1 – 3]. Existing proposals to explain the origin and the *p*-dependence of the PG could be broadly classified into two groups. One is based on the preformed pairing scenario, where PG arises from strong fluctuations of superconducting origin in the strong coupling regime for systems with low dimensionality (high structural and physical anisotropy) and low superfluid density [4]. In the second scenario, PG is attributed to some correlations of non-SC origin. Here PG is thought to coexist and, in fact, often competes with superconductivity [1]. Considerable debate has ensued as to the nature of the PG and no consensus has been reached [1 – 4]. On the other hand, the static spin/charge stripe correlations are only observed in the UD cuprates in the vicinity of $p \sim 0.125$ (the so-called 1/8$^{th}$ anomaly) [5, 6], although dynamical (fluctuating) stripe correlations are believed to exist over a much wider doping range, especially in the La214 compounds [6, 7]. Incommensurate low-energy spin fluctuations, generally interpreted as precursor to stripe correlations, are also observed in Y123 and Bi2212 [6 – 8]. Therefore, there is reason to believe that stripe ordering is a generic feature of hole doped cuprate superconductors. Stripe phase probably forms in doped Mott insulators as a compromise between the AFM ordering among the Cu spins and strong Coulomb interaction between the electrons (both favoring localization) and the kinetic energy of the mobile charge carriers (leading to delocalization). Broadly speaking, stripe phase can be viewed as spontaneously separated ordered states of charge-rich (high kinetic energy) and charge-poor (antiferromagnetically correlated) regions throughout the compound. Stripes can be both one-dimensional (1D) [6] and two-dimensional (2D) [9] in nature. It is somewhat agreed that static stripe order is detrimental to superconductivity [10], but the possible influence of fluctuating stripe order on superconductivity is a matter of significant interest [11 – 13]. Some of the theoretical models link the origins of both superconductivity and PG to the stripe correlations [11 – 13]. A proper understanding of the nature of the interplay among the SC, PG, and stripe correlations are therefore essential to construct a coherent theoretical picture explaining the normal and superconducting states of hole doped cuprates.

The study of the effects of controlled in-plane disorder on the various electronic correlations in cuprates has the potential to yield valuable information regarding their nature and possible interrelations. In a previous paper, we have reported some of the experimental features of the $p$-dependent $dT_c/dy$ for $La_{2-x}Sr_xCu_{1-y}Zn_yO_4$ only [14]. In this paper we have presented new results of the hole content dependence of the rate of suppression of $T_c$ due to Zn substitution in $YBa_2(Cu_{1-y}Zn_y)_3O_{7-\delta}$ together with those obtained for $La_{2-x}Sr_xCu_{1-y}Zn_yO_4$ [14] earlier. The observed $p$ dependences of $T_c$, $\varepsilon_g$, and $dT_c/dy$ point towards a scenario where both PG and stripe correlations are not directly related to fluctuation superconductivity. The strong stripe ordering in the vicinity of the $1/8^{th}$ anomaly affects the rate of suppression of $T_c$ in a non-trivial fashion but appears to have a minimal effect on $\varepsilon_g(p)$.

## 2. Experimental samples and measurements

A large number of polycrystalline sintered single-phase samples of $La_{2-x}Sr_xCu_{1-y}Zn_yO_4$ and $YBa_2(Cu_{1-y}Zn_y)_3O_{7-\delta}$ were synthesized by solid-state reaction method using high-purity (> 99.99%) powders. We have also used a number of high-quality crystalline $c$-axis oriented thin films of $YBa_2(Cu_{1-y}Zn_y)_3O_{7-\delta}$ grown on $SrTiO_3$ substrates by the method of pulsed laser deposition (PLD). In this paper we have used $La_{2-x}Sr_xCu_{1-y}Zn_yO_4$ samples with the following compositions (where, $x = p$): 0.08 (Zn-free only), 0.09, 0.10, 0.11, 0.12, 0.14, 0.15, 0.17 (Zn-free only), 0.19, 0.22, 0.27 (Zn-free only) and $y$ = 0.0, 0.005, 0.01, 0.015, 0.02, 0.024. The density of these sintered samples lied within the range from 5.6 gm/cm$^3$ to 6.1 gm/cm$^3$. The hole content of the Zn-substituted Y123 compounds were varied by oxygen annealing at different temperatures and gas pressures. The Y123 compounds were grown with compositions of $y$ = 0.0, 0.02, 0.03, and 0.04. All these samples were characterized by x-ray diffraction (XRD) and room-temperature thermopower ($S$[290K]). $T_c$ for sintered $La_{2-x}Sr_xCu_{1-y}Zn_yO_4$ and $YBa_2(Cu_{1-y}Zn_y)_3O_{7-\delta}$ compounds were determined from low-field ($H$ = 1 Oe, $f$ = 333.33 Hz) AC susceptibility (ACS) and resistivity, $\rho(T)$, measurements. For the epitaxial thin films of $YBa_2(Cu_{1-y}Zn_y)_3O_{7-\delta}$, $T_c$ was determined from the in-plane resistivity, $\rho_{ab}(T)$, measurements only. The details of sample preparation and characterization can be found elsewhere [15 – 17].

ACS was measured, using a commercial *Lake Shore Cryotronics* Model 7000 AC susceptometer, both on bar shaped sintered compound (with magnetic field applied along the longest dimension) and on powder. The AC susceptometer was calibrated using a Pb sphere. When completely diamagnetic, such a sphere yields a signal of 23.9 $\mu V/mm^3$ (with $H$ = 1 Oe, $f$ = 333.33 Hz). Most of the samples used in this study exhibited sharp SC transitions ($\Delta T_c$ < 0.5 K). Transition widths increase somewhat for the heavily Zn substituted compounds. ACS results for some of the sintered $La_{2-x}Sr_xCu_{1-y}Zn_yO_4$ compounds are shown in Figs. 1. $T_c$ was determined keeping the resolution of the susceptometer and the effect of Gaussian fluctuation diamagnetism in mind as follows: a straight line was drawn at the steepest part near the onset of the diamagnetic ACS curve (where the magnitude of the ACS signal does not exceed a few microvolts, this follows from the estimated orders of magnitude of the contribution to the ACS originating from diamagnetic fluctuations just above $T_c$ for La214 and Y123 [18]), and another one was drawn as the $T$-independent base line associated with negligibly small normal state signal. The intercept of the two lines gave $T_c$. A typical example is shown in Fig. 2. This procedure yields almost identical (~ 1 K higher for the powdered compounds) values of $T_c$ both for sintered and powdered samples. Resistively, $T_c$ was taken as the average of the zero-resistivity temperature obtained from both the cooling and warming runs. Resistivity was measured using the four-terminal configuration. The thin films were patterned accordingly. $T_c$ for the thin films were ~ 1 K lower than those obtained for the sintered $YBa_2(Cu_{1-y}Zn_y)_3O_{7-\delta}$ compounds with identical $p$, possibly due to epitaxial strain. $T_c$ values obtained from resistivity and ACS measurements agree quite well, within 1 K, where applicable. Representative resistivity data are plotted in Figs. 3. The values of the hole content was obtained from the values of $S$[290K] [19] which is quite insensitive to the crystalline state as well as the disorder content. Iso-valent Zn substitution does not alter the hole content in the $CuO_2$ plane significantly. The $T_c(p, y)$ values obtained from ACS and resistivity data agree quite well, where available, with those found in previous studies [20, 21].

We have plotted the $T_c(p, y = 0)$ and $dT_c(p)/dy$ results for $La_{2-x}Sr_xCu_{1-y}Zn_yO_4$ and $YBa_2(Cu_{1-y}Zn_y)_3O_{7-\delta}$ in Figs. 4a and 4b respectively. For direct comparison we have shown $dT_c(p)/dy$ for $La_{2-x}Sr_xCu_{1-y}Zn_yO_4$ and $YBa_2(Cu_{1-y}Zn_y)_3O_{7-\delta}$ together in Fig. 5. Figs. 4a and 4b show clearly the strongly $p$-dependent nature of $dT_c(p)/dy$. It is seen that the magnitude of $dT_c(p)/dy$, except near $p \sim 0.125$, decreases systematically with increasing $p$ in the UD to optimally doped region (passing through a minimum in the OD near $p \sim 0.20$ and increases slowly again for further overdoping for $La_{2-x}Sr_xCu_{1-y}Zn_yO_4$). Size of this anomaly is significantly less pronounced for the $YBa_2(Cu_{1-y}Zn_y)_3O_{7-\delta}$ compounds. In some recent work we have extracted the PG energy scale from the analysis of the uniform magnetic susceptibility data for some of the compounds used in this study, details of which can be found in refs. [22 – 24]. We show those $\varepsilon_g(p)$ values in Figs. 6 for $La_{2-x}Sr_xCu_{1-y}Zn_yO_4$ and $YBa_2(Cu_{1-y}Zn_y)_3O_{7-\delta}$. The $\varepsilon_g(p)$ values shown here are in excellent agreement with those found by other studies using different experimental probes [1]. It is worth noticing that $\varepsilon_g(p)$ appears almost featureless at $p \sim 0.12$ (Fig. 6).

## 3. Discussion and conclusions

The anomalous decrease of $dT_c(p)/dy$ in the vicinity of the 1/8$^{th}$ doping imply that the effect of Zn on degrading $T_c(p)$ becomes less effective when stripe ordering is at its strongest. In general, the non-magnetic and iso-valent Zn induced rate of suppression of $T_c$ in cuprates can be reasonably well-understood within the framework of strong potential (unitary) scattering of Cooper pairs with $dx^2-y^2$ order parameter. Evidence for such pair-breaking scattering by Zn was clearly found from the phase shift analysis of the pioneering STM results obtained by Pan *et al.* [25]. In this unitary scattering formalism, the *p*-dependent variation of $dT_c/dy$ mainly arises from the *p*-dependence of the PG energy scale [26, 27], as the scattering rate is inversely proportional to the thermal average of the electronic density of states (EDOS) at the Fermi-level. Such a picture can address the observed behavior of the $dT_c(p)/dy$ in the UD $La_{2-x}Sr_xCu_{1-y}Zn_yO_4$ and $YBa_2(Cu_{1-y}Zn_y)_3O_{7-\delta}$ and slightly OD compounds except around the 1/8$^{th}$ anomaly. The question that naturally arises, is what happens to the PG energy scale (which roughly measures the degree of depletion of the EDOS at Fermi-level) at $p \sim 0.125$? Since $\varepsilon_g(p)$ varies monotonically around this hole concentration, it is reasonable to assume that the anomaly in $dT_c/dy$ is not directly related to the QP spectral density. The minimum in the magnitude of $dT_c(p)/dy$ at $p \sim 0.19$ for $La_{2-x}Sr_xCu_{1-y}Zn_yO_4$ follows naturally as the PG disappears at this hole content [1, 20, 22, 28, 29]. Such a minimum in $dT_c/dy$ at $p \sim 0.19$ was also observed in Ca substituted OD Y123 compounds [27].

On the hand, the slow increment in the magnitude of $dT_c(p)/dy$ in the OD side above $p \sim 0.19$, is indicative of a fundamental change in the electronic ground state of different origin [1, 29] where the 3D [26] Fermi-liquid description becomes more applicable and a tendency towards metal-superconducting phase separation starts to grow [30]. It is important to note that the above description supports a non-SC origin for the PG. Here PG competes with superconductivity, at least in the sense that it takes away QP states which otherwise would have been available for the SC condensate (adding to the superfluid density). Also, the fact that Zn degrades $T_c$ rapidly but does not affect the PG energy scale [20, 21, 28] lends further support for the non-SC origin of the PG.

The size of the anomaly observed in $dT_c(p)/dy$ near $p \sim 0.125$ indicates that a different mechanism is at play, affecting the pair-breaking. Considering that static stripe order reduces $T_c$ and the superfluid density is also suppressed in the vicinity of $p \sim 0.12$ [31], it is rather surprising to find Zn becoming less effective in reducing $T_c$ in samples where superconductivity is already weakened. This observation becomes even more counterintuitive when one takes into account of the Zn induced pinning of the stripe fluctuations [32 – 34]. The pinning or slowing down of spin fluctuations by Zn can be attributed to the enhancement of the AFM correlations, carrier localization, or to the increase in the stripe inertia. Irrespective of the precise mechanism, as static charge/spin ordering degrades superconductivity, Zn substitution should become more effective in reducing $T_c$ near the $1/8^{th}$ doping. Exactly the opposite effect is found experimentally. The possible reasons for this can be the followings (i) since spin/charge ordering near $p \sim 0.125$ is already static or quasi-static in the pure compound, Zn substitution plays no significant role in further pinning in this region. In this scenario Zn substitution in cuprates with $p$ close to the $1/8^{th}$ value is "*wasted*" to some extent. This supports the theoretical proposal by Smith *et al*. [35] which describes stripe-pinning as primary mechanism for degradation of $T_c$ due to Zn. (ii) Zn substitution destroys the integrity of the static stripe order. For randomly substituted Zn atoms in the Cu sites, part of the Zn in the hole-rich regions will lead to carrier localization and the other part will replace the antiferromagnetically correlated Cu spins in the hole-poor regions, thereby creating spin vacancies. In such situation large number of spin vacancies, a hole from a neighboring domain can hop inside the spin ordered region. This process will destroy the stripe order itself [36]. Within this proposal Zn becomes less effective in reducing $T_c$ because the static stripe order is weakened and consequently superconductivity itself somewhat *strengthened*. The tendency towards stripe formation is stronger in La214 compounds and weaker in Y123 [6, 7]. Incommensurate low-energy spin fluctuations are present in Y123 over a wide range of $p$ [6] but there is no concrete evidence of charge ordering in these materials. This offers an explanation for the reduced size of the anomaly in $dT_c(p)/dy$ around $p = 0.125$ in YBa$_2$(Cu$_{1-y}$Zn$_y$)$_3$O$_{7-\delta}$.

As mentioned in previous sections, absence of any noticeable feature in $\varepsilon_g(p)$ near the $1/8^{th}$ doping indicates that PG and stripes might be unrelated phenomena. The normal state Nernst signal in cuprate superconductors [37] are believed to be arising from fluctuating Cooper pairs and a number of papers have linked it to the existence of the PG [37 – 39]. The Nernst signal increases significantly in La214 compounds in the vicinity of the $1/8^{th}$ doing [40]. Recent experimental and theoretical studies [41, 42] have linked this enhancement to a reconstruction (translational symmetry breaking) of the Fermi surface. For systems with low Fermi energy a large Nernst effect can result solely from the appearance of electron or hole pockets [43], and PG and stripe correlations can remain unrelated. It should be mentioned that recently Parker *et al.* [44] have proposed a scenario, based on the STM experiments on Bi2212, where the existence of the fluctuating stripes were taken as a consequence of the PG rather than the other way around. This interesting proposal calls for further experimental and theoretical investigations.

To conclude, we have investigated the effect of Zn on $T_c$ as a function of in-plane hole concentration for two different families of cuprates: single-plane $La_{2-x}Sr_xCu_{1-y}Zn_yO_4$ and double-plane $YBa_2(Cu_{1-y}Zn_y)_3O_{7-\delta}$, over a wide range of compositions. A systematic variation in $dT_c(p)/dy$ was found for both families except near $p \sim 0.125$. The anomaly around $1/8^{th}$ doping is lower for the Y123 compounds. The characteristic $\varepsilon_g(p)$ energy scale is featureless at this doping and shows no significant difference for La214 and Y123 over the entire experimental range of $p$. We have discussed the possible scenarios for this anomaly in the magnitude of $dT_c(p)/dy$ close to the $1/8^{th}$ doping. Contrasting effects of Zn on $T_c$, $\varepsilon_g(p)$, and stripe dynamics indicate that they are probably unrelated phenomena. A similar conclusion was drawn by Tallon *et al*. [45] from their oxygen isotope exponent measurements on various hole doped cuprate superconductors.

## Acknowledgements

The authors thank Prof. J. R. Cooper and Dr. J. W. Loram, University of Cambridge, UK, and Prof. J. L. Tallon, Industrial Research Limited, Wellington, New Zealand, for useful discussions at various stages. The authors also acknowledge the Commonwealth Commission, UK, and Trinity College, University of Cambridge, UK, for financial support and support with experimental facilities. SHN would also like to thank the AS-ICTP, Trieste, Italy, for the hospitality, where writing-up of this paper was done.

## References

[1] Tallon J L and Loram J W 2001 *Physica C* **349** 53
[2] Lee P A 2008 *Rep. Prog. Phys.* **71** 012501
[3] Damascelli A, Hussain Z and Shen Z -X 2003 *Rev. Mod. Phys.* **75** 473
[4] Emery V J and Kivelson S A 1995 *Nature* **374** 434
[5] Tranquada J M, Sternlib B J, Axe J D, Nakamura Y and Uchida S 1995 *Nature* **375** 561
[6] Berg E, Fradkin E, Kivelson S A and Tranquada J M 2009 *New J. Phys.* **11** 115004
[7] Kivelson S A, Bindloss I P, Fradkin E, Oganesyan V, Tranquada J M, Kapitulnik A and Howard C 2003 *Rev. Mod. Phys.* **75** 1201

[8] Parker C V, Aynajian P, da Silva Neto E H, Pusp A, Ono S, Wen J, Xu Z, Gu G and Yazdani A 2010 *Nature* **468** 677

[9] Hinkov V, Pailhes S, Bourges P, Sidis Y, Ivanov A, Kulakov A, Lin C T, Chen D P, Bernhard C and Keimer B 2004 *Nature* **430** 650

[10] Carlson E W, Emery V J, Kivelson S A and Orgad D in *The Physics of Superconductors Vol. II. Superconductivity in Nanostructure, High-Tc and Novel Superconductors, Organic Superconductors,* Eds. Bennemann K H, Ketterson J B, *Springer-Verlag* 2004

[11] Castellani C, Castro C and Di Grilli M 1995 *Phys. Rev. Lett.* **75** 4650

[12] Tranquada J M 2005 *J. Phys.* **IV** *France* 1

[13] Granath M and Anderson B M 2010 *Phys. Rev. B* **81** 024501

[14] Islam R S and Naqib S H 2008 *Supercond. Sci. Technol.* **21** 125020

[15] Naqib S H 2003 *Ph.D. thesis*, University of Cambridge (unpublished)

[16] Naqib S H, Chakalov R A and Cooper J R 2004 *Physica C* **407** 73

[17] Islam R S 2005 *Ph.D. thesis*, University of Cambridge (unpublished)

[18] Bulaevskii L N, Ginzburg V L and Sobayanin A A 1988 *Zh. Eksp. Teor. Fiz.* **94** 355

[19] Obertelli S D, Cooper J R and Tallon J L 1992 *Phys. Rev.* B **46** 14928 and Tallon J L, Cooper J R, DeSilva P S I P N, Williams G V M and Loram, J W 1995 *Phys. Rev. Lett.* **75** 4114

[20] Naqib S H, Cooper J R, Tallon J L and Panagopoulos C 2003 *Physica* C **387** 365

[21] Walker D J C, Mackenzie A P and Cooper J R 1995 *Phys. Rev. B* **51** 15653

[22] Naqib S H and Islam R S 2008 *Supercond. Sci. Technol.* **21** 105017

[23] Islam R S and Naqib S H 2010 Physica C **470** 79

[24] Islam R S, Hasan M M and Naqib S H 2010 *J. Supercond. Nov. Magn*. **23** 1569

[25] Pan S H, Hudson E W, Lang K M, Eisaki H, Uchida S and Davis J C 2000 *Nature* **403** 746

[26] Tallon J L, Bernhard C, Williams G V M and Loram J W 1997 *Phys. Rev. Lett*. **79**, 5294

[27] Naqib S H 2007 *Supercond. Sci. Technol*. **20**, 964

[28] Naqib S H, Cooper J R, Tallon J L, Islam R S and Chakalov R A 2005 *Phys. Rev.* B **71** 054502

[29] Rourke P M C, Mouzopoulo I, Xu X, Panagopoulos C, Wang Y, Vignolle B, Proust C, Kurganova E V, Zeitler U, Tanabe Y, Adachi T, Koike Y and Hussey N E 2011 *Nature Physics, doi: 10.1038/nphys 1945*

[30] Wen H H, Chen X H, W. Yang W L and Zhao Z X 2000 *Phys. Rev. Lett*. **85** 2805

[31] Panagopoulos C and Dobrosavljevic V 2005 *Phys. Rev. B* **72** 014536

[32] Akoshima M, Noji T, Ono Y and Koike Y 1998 *Phys. Rev. B* **57** 7491

[33] Akoshima M, Koike Y, Watanabe I and Nagamine K 2000 *Phys. Rev. B* **62** 6721

[34] Risdiana, Adachi T, Oki N, Yairi S, Tanabe Y, Omori K, Koike Y, Suzuki T, Watanabe I, Koda A and Higemoto W 2008 *Phys. Rev.* B **77** 054516

[35] Smith C M, Castro Neto A H and Balatsky A V 2000 *cond-mat/0012080* (unpublished)

[36] Anegawa O, Okajima Y, Tanda S and Yamaya K 2001 *Phys. Rev.* B **63** 140506

[37] Xu Z A, Ong N P, Wang Y, Kakeshita T and Uchida S 2000 Nature **406** 486

[38] Wang Y, Li L and Ong N P 2006 *Phys. Rev. B* **73** 024510

[39] Wang Y, Li L, Naughton M J, Su G D, Uchida S and Ong N P 2005 *Phys. Rev. Lett*. **95** 247002

[40] Cyr-Choinere O, Daou R, Laliberte F, LeBoeuf D, Doiron-Leyraud N, Chang J, Yang J -Q, Cheng J -G, Zhou J -S, Goodenough J B, Pyon S, Takayama T, Takagi H, Tanaka Y and Taillefer L 2009 *Nature* **458** 743

[41] Laliberte L, Chang J, Doiron-Leyraud N, Hassinger E, Daou R, Rondeau M, Ramshaw B J, Liang R, Bonn D A, Hardy W N, Pyon S, Takayama T, Takagi H, Sheiken I, Malone L, Proust C, Behnia K and Taillefer L 2011 *cond-mat/1102.0984* (unpublished)

[42] Hackl A, Vojta M and Sachdev S 2010 *Phys. Rev. B* **81** 045102

[43] K. Behnia 2009 *J. Phys. Cond. Matt*. **21** 113101

[44] Colin V Parker, Pegor Aynajian, Eduardo H da Silva Neto, Akash Pusp, Shimpeio Ono, Jinseng Wen, Zhijun Xu, Genda Gu and Ali Yazdani 2010 *Nature* **468** 677.

[45] Tallon J L, Islam R S, Storey J, Williams G V M and Cooper J R 2005 *Phys. Rev. Lett.* **94** 237002

[46] Presland M R, Tallon J L, Buckley R G, Liu R S and Flower N E 1991 *Physica C* **176** 95

**Figure captions**

Figure 1 (color online): ACS data for some representative sintered $La_{2-x}Sr_xCu_{1-y}Zn_yO_4$ compounds [14] with (a) $x = 0.19$, (b) $x = 0.09$, and (c) $x = 0.11$ (close to $1/8^{th}$ anomaly). Zn contents (*y*-values) are shown in the plots.

Figure 2 (color online): Extraction of $T_c$ from the ACS plots (see text for details) for $La_{1.81}Sr_{0.19}Cu_{1-y}Zn_yO_4$ [14] (*y*-values are shown in the plot).

Figure 3 (color online): In-plane resistivity, $\rho_{ab}(T)$, for some representative *c*-axis oriented crystalline thin films of $YBa_2(Cu_{1-y}Zn_y)_3O_{7-\delta}$ with (a) $y = 0.00$, (b) $y = 0.02$, (c) $y = 0.04$, and (d) Resistivity data for $y = 0.03$ sintered compounds. The hole contents (within $\pm$ 0.004) are given in the plots.

Figure 4 (color online): $T_c(y = 0)$ and $dT_c/dy$ versus $p$ for (a) $La_{2-x}Sr_xCu_{1-y}Zn_yO_4$ and (b) $YBa_2(Cu_{1-y}Zn_y)_3O_{7-\delta}$. The thick dotted/dashed lines are drawn as guides to the eyes.

Figure 5 (color online): $dT_c/dy$ versus $p$ for $La_{2-x}Sr_xCu_{1-y}Zn_yO_4$ and $YBa_2(Cu_{1-y}Zn_y)_3O_{7-\delta}$ shown together. The boxed part encloses the anomaly near the $1/8^{th}$ doping.

Figure 6 (color online): The characteristic PG energy scales, $\varepsilon_g(p)$, expressed in temperature, for $La_{2-x}Sr_xCu_{1-y}Zn_yO_4$ and $YBa_2(Cu_{1-y}Zn_y)_3O_{7-\delta}$. The filled symbols denote Zn-free compounds and the open symbols are for the $y = 0.02$ La214 compounds. The thick straight line shows almost a linear decrease in $\varepsilon_g$ as $p$ increases. The broken lines show parabolic $T_c(p)$ relation [46] using maximum $T_c$ values as 39 K and 92 K at optimum doping for La214 and Y123, respectively.

Figure 1

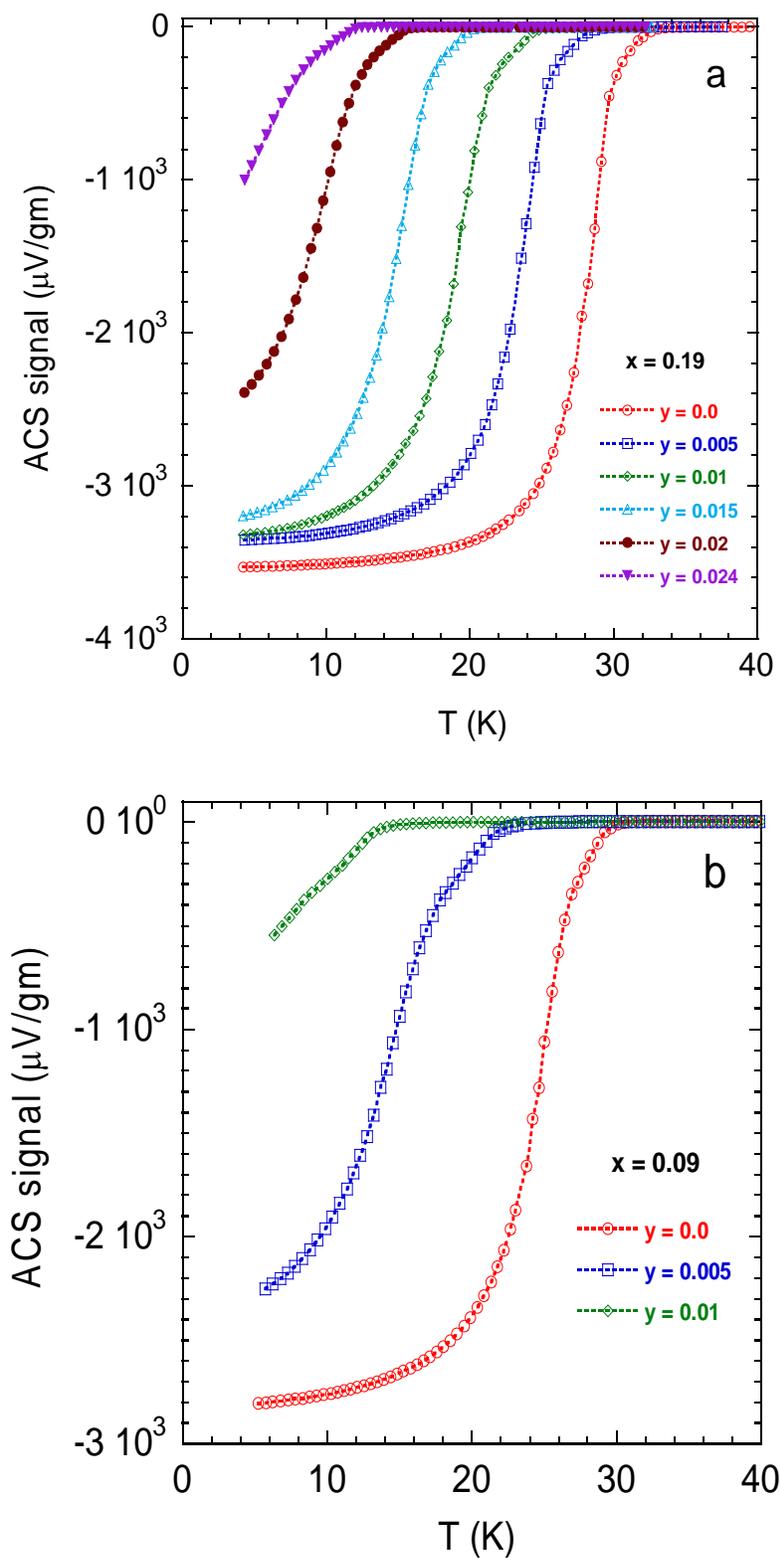

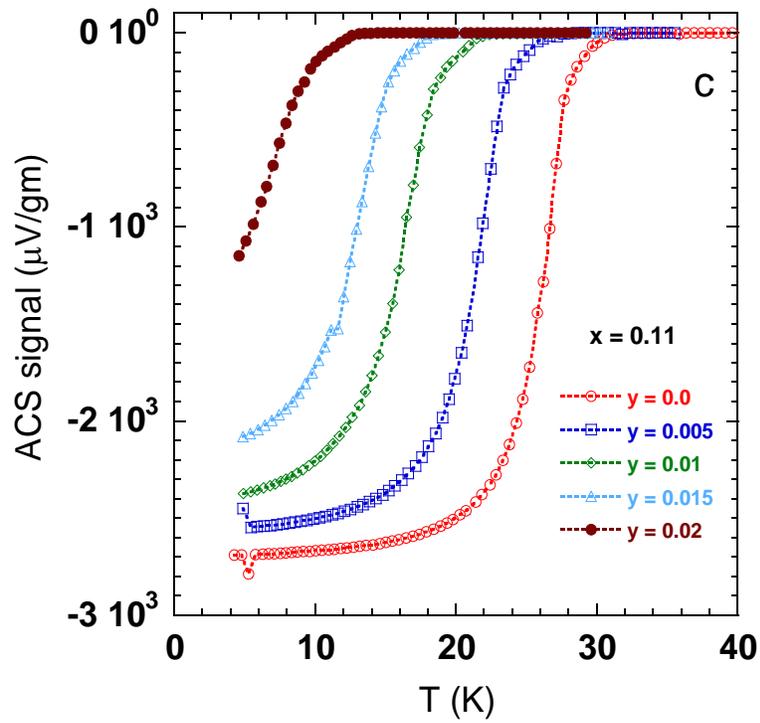

Figure 2

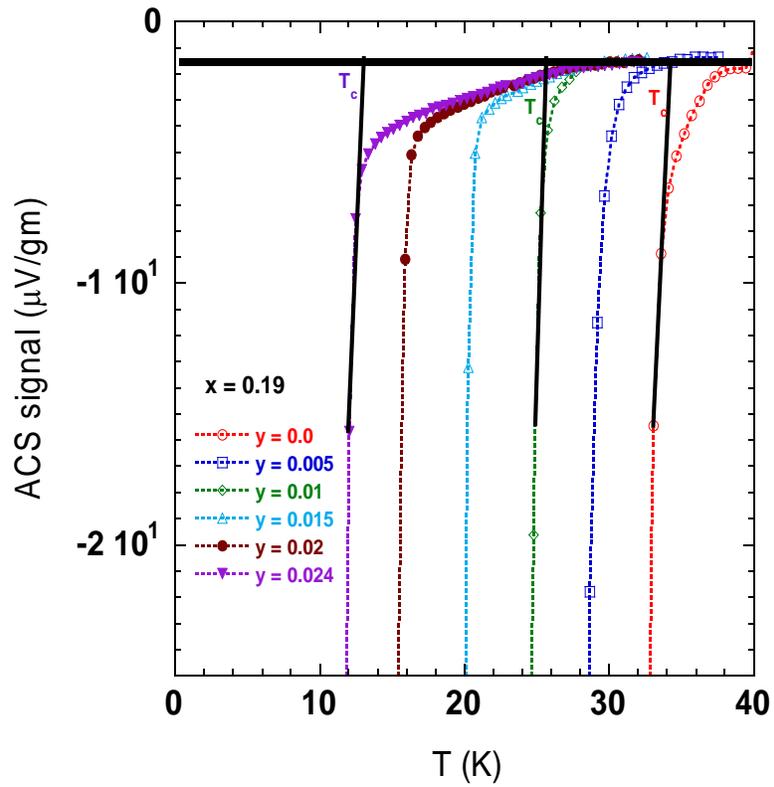

Figure 3

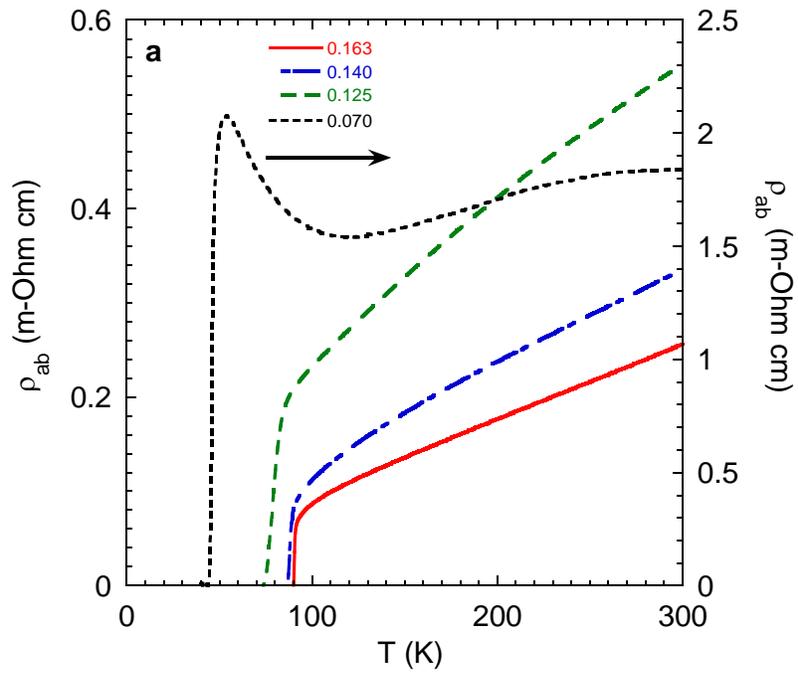

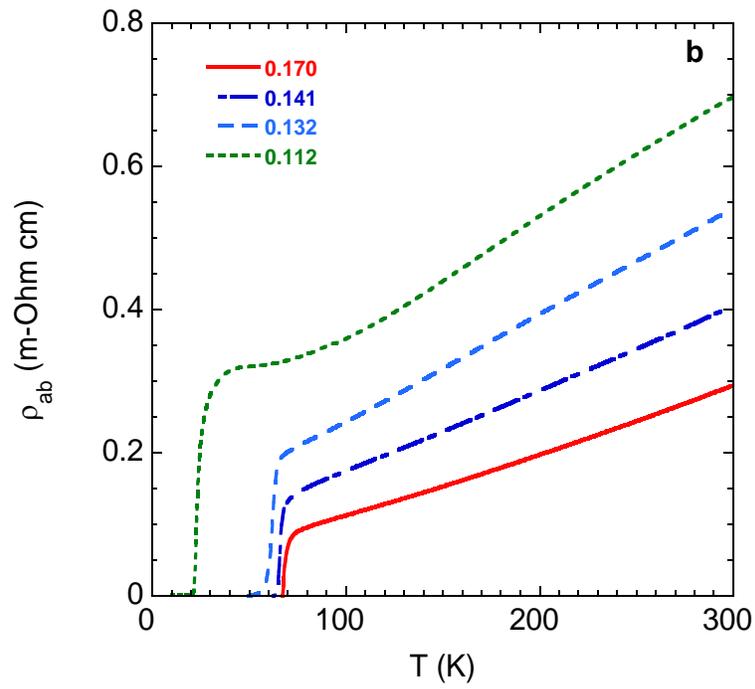

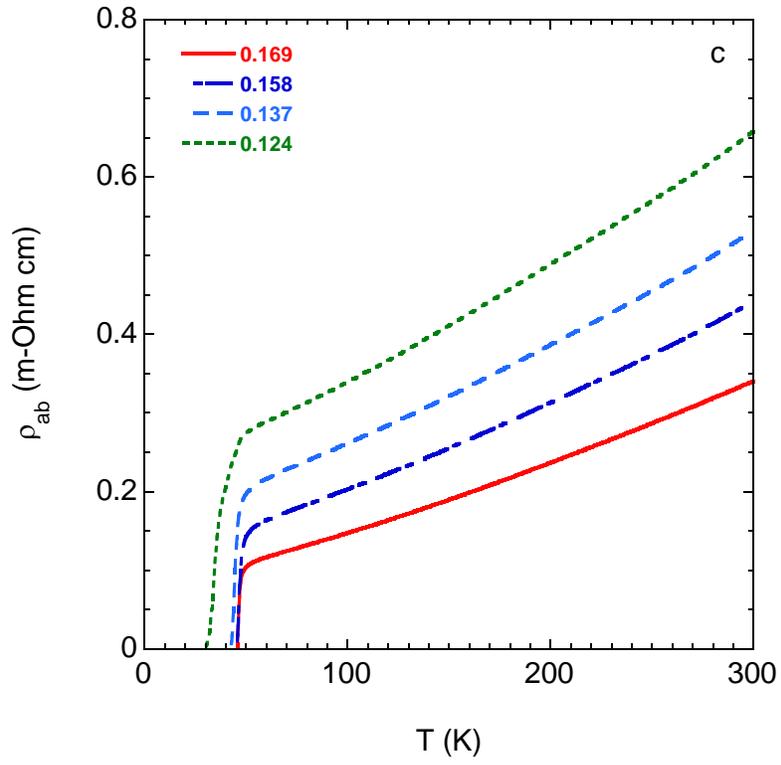

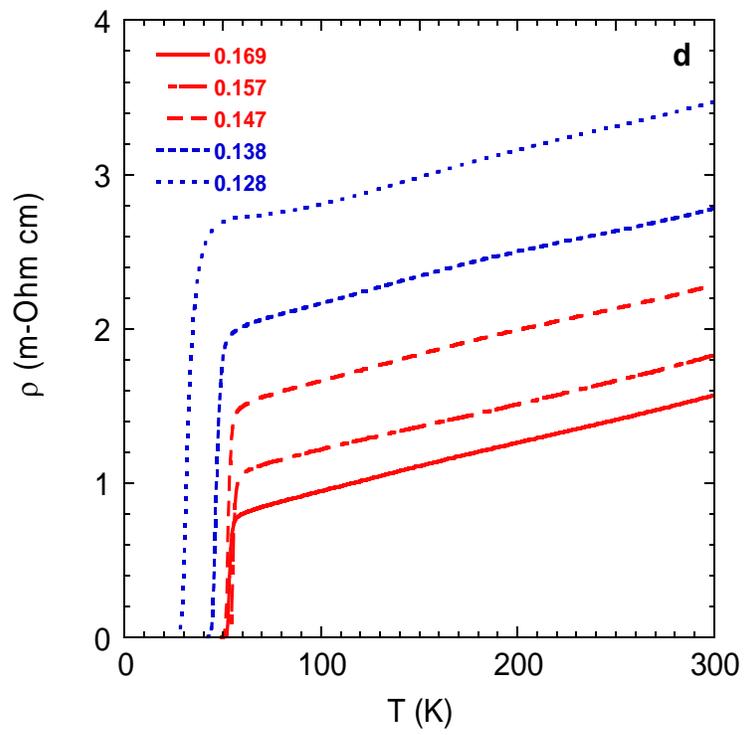

Figure 4

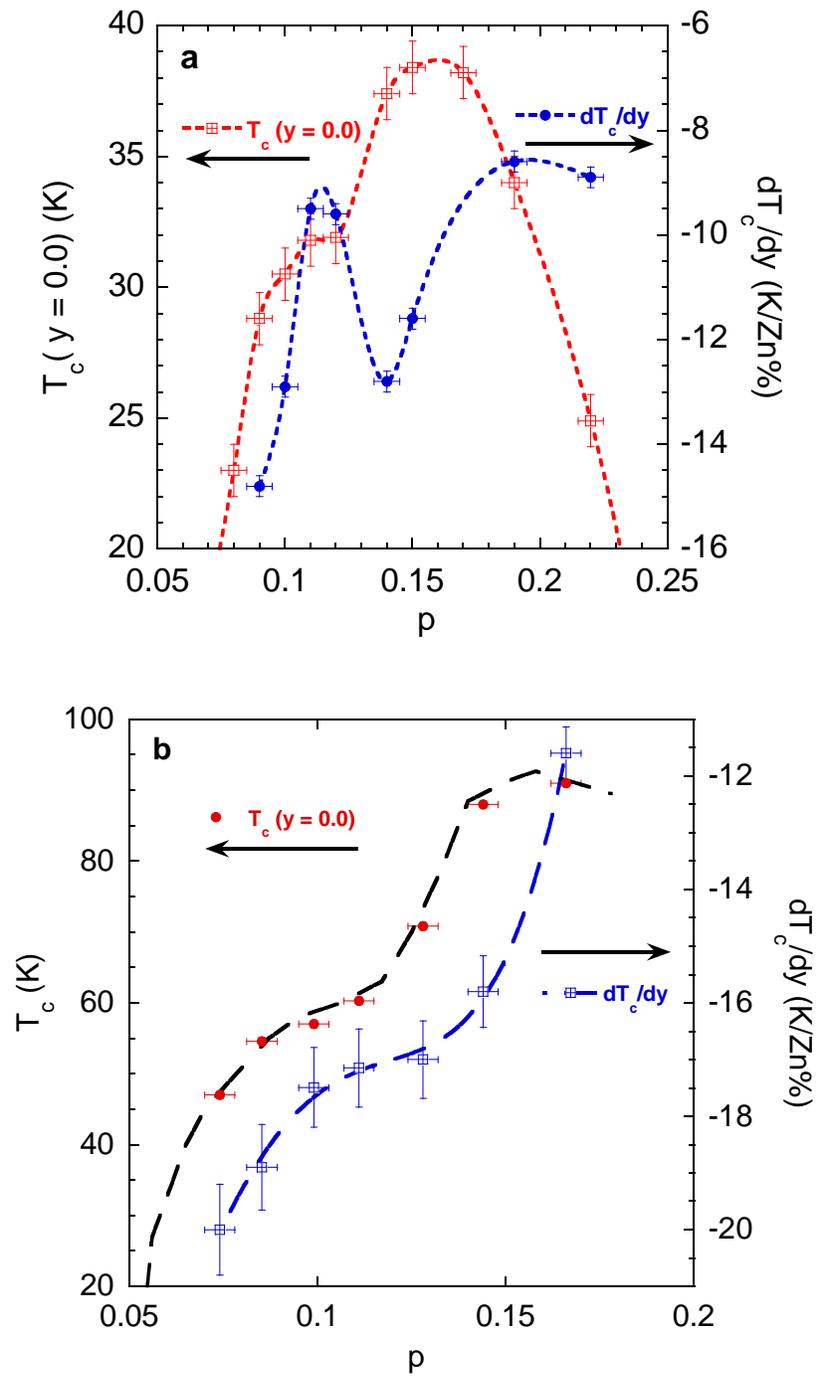

Figure 5

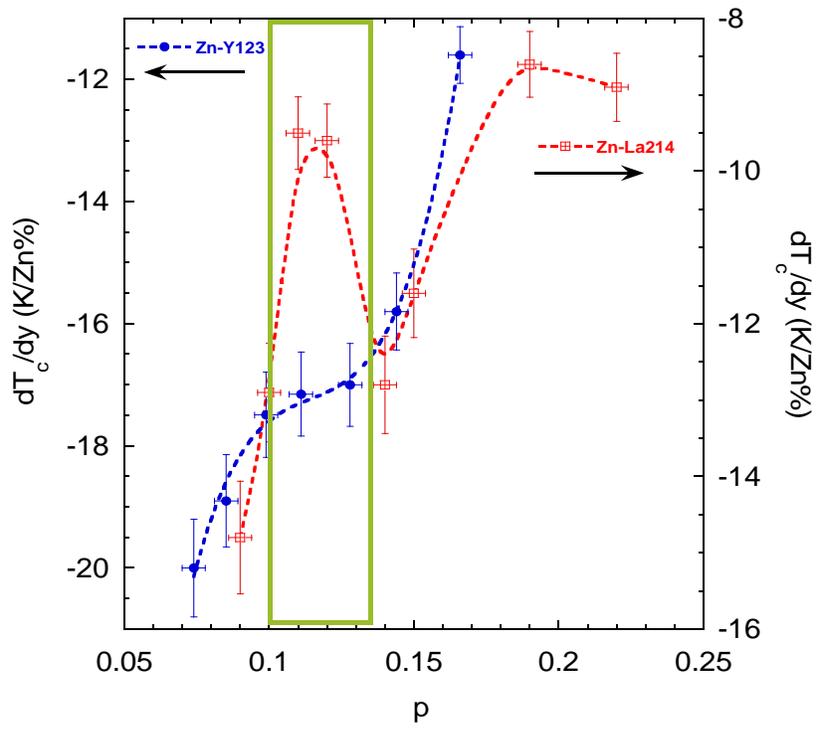

Figure 6

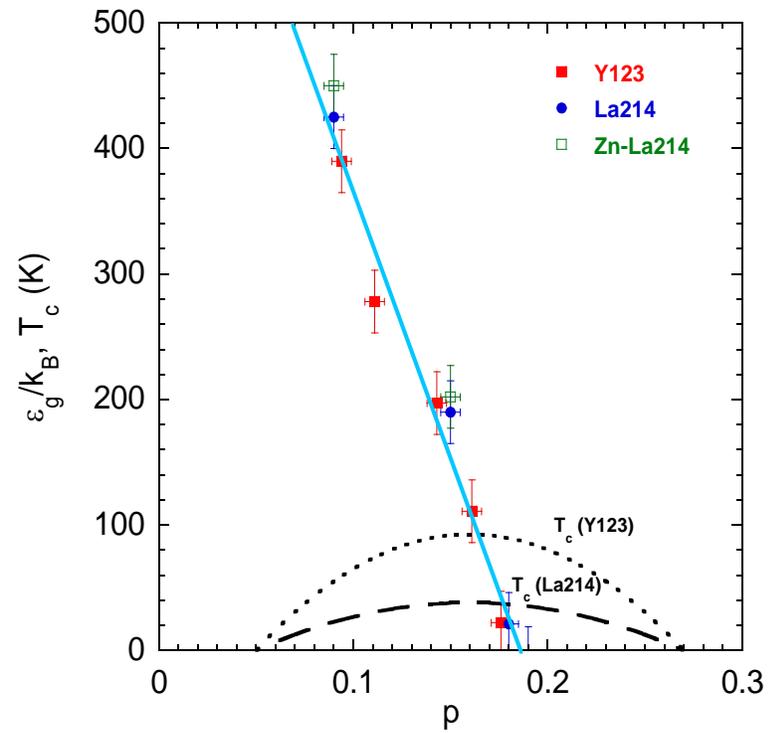

16